\pdfoutput=1
\documentclass[final,5p,11pt,authoryear,hidelinks,times]{elsarticle}
\usepackage[final]{microtype}
\usepackage{setspace}

\makeatletter
\let\@afterindenttrue\@afterindentfalse
\makeatother

\usepackage{titlesec}
\titleformat*{\section}{\Large\normalshape\bf}
\titleformat*{\subsection}{\large\normalshape\bf}
\titleformat*{\subsubsection}{\normalshape\bf}

\makeatletter
\renewcommand\paragraph{\@startsection{paragraph}{4}{\z@}%
  {-3.25ex\@plus -1ex \@minus -.2ex}%
  {1.5ex \@plus .2ex}%
  {\normalfont\normalsize\bfseries\itshape}}
\makeatother
\journal{Computational Biomechanics for Medicine}

\usepackage{hyperref}

\usepackage[font=normalsize,labelfont=bf]{caption}

\usepackage[utf8]{inputenc}

\usepackage{float}

\usepackage{amsmath}
\usepackage{amsfonts}
\usepackage{bm}
\usepackage{siunitx}

\usepackage{tabularx}
\usepackage{booktabs}

\usepackage{xcolor}

\usepackage{graphicx}
\usepackage{subcaption}

\usepackage[status=final,index,multiuser,author=]{fixme}
\fxsetface{margin}{\linespread{1}\scriptsize}
\fxusetheme{color}

\usepackage[displaymath]{lineno}
\newcommand*\patchAmsMathEnvironmentForLineno[1]{%
  \expandafter\let\csname old#1\expandafter\endcsname\csname #1\endcsname
  \expandafter\let\csname oldend#1\expandafter\endcsname\csname end#1\endcsname
  \renewenvironment{#1}%
     {\linenomath\csname old#1\endcsname}%
     {\csname oldend#1\endcsname\endlinenomath}}%
\newcommand*\patchBothAmsMathEnvironmentsForLineno[1]{%
  \patchAmsMathEnvironmentForLineno{#1}%
  \patchAmsMathEnvironmentForLineno{#1*}}%
\AtBeginDocument{%
\patchBothAmsMathEnvironmentsForLineno{equation}%
\patchBothAmsMathEnvironmentsForLineno{align}%
\patchBothAmsMathEnvironmentsForLineno{flalign}%
\patchBothAmsMathEnvironmentsForLineno{alignat}%
\patchBothAmsMathEnvironmentsForLineno{gather}%
\patchBothAmsMathEnvironmentsForLineno{multline}%
}

\begin{document}

\begin{frontmatter}

  \title{
    Open Meshed Anatomy: Towards a comprehensive \\ finite element hexahedral mesh derived from open atlases}

\address[UWA]{
  Intelligent Systems for Medicine Laboratory,
  The University of Western Australia,\\
  35 Stirling Highway,
  Perth, WA, Australia}

\address[SPL]{
  Surgical Planning Laboratory,
  Department of Radiology of Brigham and Women's Hospital,
  Boston, MA, USA}

\address[Harvard]{
  Harvard Medical School,
  Boston, MA, USA}

\author[UWA]{A.T. Huynh\corref{mycorrespondingauthor}}
\ead{andy.huynh@research.uwa.edu.au}
\cortext[mycorrespondingauthor]{Corresponding author}

\author[UWA]{B.F. Zwick}

\author[SPL,Harvard]{M. Halle}

\author[UWA]{A. Wittek}

\author[UWA,Harvard]{K. Miller}

\begin{abstract}

Computational simulations using methods such as the finite element (FE) method rely on high-quality meshes for achieving accurate results. This study introduces a method for creating a high-quality hexahedral mesh using the Open Anatomy Project's brain atlas. Our atlas-based FE hexahedral mesh of the brain mitigates potential inaccuracies and uncertainties due to segmentation — a process that often requires input of an inexperienced analyst. It accomplishes this by leveraging existing segmentation from the atlas. We further extend the mesh's usability by forming a two-way correspondence between the atlas and mesh. This feature facilitates property assignment for computational simulations and enhances result analysis within an anatomical context. We demonstrate the application of the mesh by solving the electroencephalography (EEG) forward problem. Our method simplifies the mesh creation process, reducing time and effort, and provides a more comprehensive and contextually enriched visualisation of simulation outcomes.

\end{abstract}


\end{frontmatter}

\section{Introduction}
Computational simulations using methods such as the finite element (FE) method rely on high-quality meshes for achieving accurate results. High-quality meshes can be particularly challenging to produce for intricate structures and organs found in the human body. The human brain is an extremely complex organ because of its structure and sophisticated neuroanatomy. The creation of FE meshes for brain simulations requires substantial effort and often entails a time-intensive process. It is often observed that novice researchers in this field may dedicate many months of their efforts in learning and constructing patient-specific brain models prior to initiating any simulations. This is likely due to the potential lack of sufficient medical knowledge, preventing them from accurately identifying various human anatomical structures and executing manual segmentation to isolate their regions of interest. Moreover, challenges persist in the field with regard to the automatic generation of high-quality meshes, particularly hexahedral meshes. The optimisation of mesh element quality necessitates not only significant experience but also manual intervention.

Engineers from Sandia National Laboratories report that creating the FE model takes up approximately 80\% of the total analysis time, with only 20\% being devoted to the actual analysis \citep{Cottrell_2009}. Based on our experience, creating a patient-specific model with an acceptable quality should take no more than 40 minutes using standard computing hardware and a basic level of domain knowledge \citep{Wittek_2016}. This requirement is based on the need for compatibility with clinical workflows, as it is unrealistic to expect hospitals to possess expertise in numerical analysis or have access to supercomputing resources in practical settings. Streamlining the process to create computational meshes is essential for advancing research and translating these methods into clinical applications.

Another aspect of computational simulation that needs improvement is the visualisation of simulation results. Currently, simulation results of the human brain are often mapped onto 3D medical images or 3D meshes offering limited anatomical information and limiting context for detailed analysis. For example, the displacement field vectors resulting from simulations of brain shift are typically projected onto grayscale MRI images \citep{Safdar_2023}. This approach makes interpreting the affected regions challenging. Understanding these displacements in relation to specific anatomical structures and their impact is a vital consideration not fully addressed by current methods. Even experienced medical professionals may struggle with interpretation of these computational results due to the absence of explicit anatomical structure labels.

Improvement is needed in the areas of mesh creation and visualisation of results for computational simulations before they can transition into medical applications. Having an atlas-based FE mesh open to the public significantly reduces the time and effort required for researchers to create their own models. Additionally, improving visualisation methods for simulation outcomes could offer a more comprehensive analysis, benefiting researchers and clinicians alike.

\subsection{Objectives and Scope}

In this study, we present a method for creating a high-quality hexahedral mesh using the Open Anatomy Project's SPL/NAC brain atlas. This atlas was released by the Neuroimage Analysis Center’s (NAC’s) Computational Clinical Anatomy Core and the Surgical Planning Laboratory (SPL) at Brigham and Women’s Hospital in 2017 and comprises of over 300 anatomical structures from a healthy individual's brain \citep{Halle_2017}. With development extending over two decades, this comprehensive atlas offers significant advantages for computational simulations.

One advantage lies in the atlas’ high quality, labelled map of segmented brain structures. This eliminates the need for manual segmentation of individual subjects or patients, thereby reducing potential inaccuracies due to over or under-segmentation by inexperienced users. By commencing with the comprehensive and accurately segmented SPL/NAC brain atlas, we can generate a highly detailed hexahedral mesh directly from the labelled medical images. Such an approach is beneficial by utilising methods such as medical image registration to warp the mesh onto different subjects or patient, providing a more efficient means of creating patient-specific models.

Another significant advantage underlined in our research is the formation of a two-way correspondence between the atlas and the mesh. This method improves the functionality of the mesh by annotating elements or nodes with the respective anatomical brain structures. Such a system can be favourable for assigning properties for simulations and cross-examining simulation outcomes in an anatomical context, instead of limitations brought about with spatial coordinates alone.

We outline the method for creating this atlas-based mesh using extensions and modules in the open-source medical image analysis and visualisation software, 3D Slicer \citep{Fedorov_2012}. We have utilised this atlas-based mesh by solving a simplified electrostatic problem based on our recent publications \citep{Safdar_2023}, \citep{Zwick_2022}. We employ the SlicerCBM extension to predict the electric field initiated by a dipole located within the brain. SlicerCBM is an automatic framework for the biophysical analysis of the brain developed by our team \citep{Safdar_2023}.

\section{Methods}
\subsection{Data acquisition and processing}

The SPL/NAC brain atlas data is of a healthy 42-year-old male volunteer, acquired at the Martinos Center for Biomedical Imaging on a Siemens 3T scanner, using a multi-array head coil. Labels were generated using Freesurfer’s automatic parcellation followed by manual segmentation of many structures. The dataset includes a volumetric whole head MPRAGE and T2-weighted series with a voxel size of 0.75 mm isotropic, a downscaled version of both acquisitions at 1 mm isotropic resolution, a per-voxel labelling of structures based on the 1 mm volumes, and a hierarchy of structures loosely based on the Radlex ontology, which is provided in a JSON file \citep{Halle_2017}.

\begin{figure*}
    \centering
    \includegraphics[width=\textwidth]{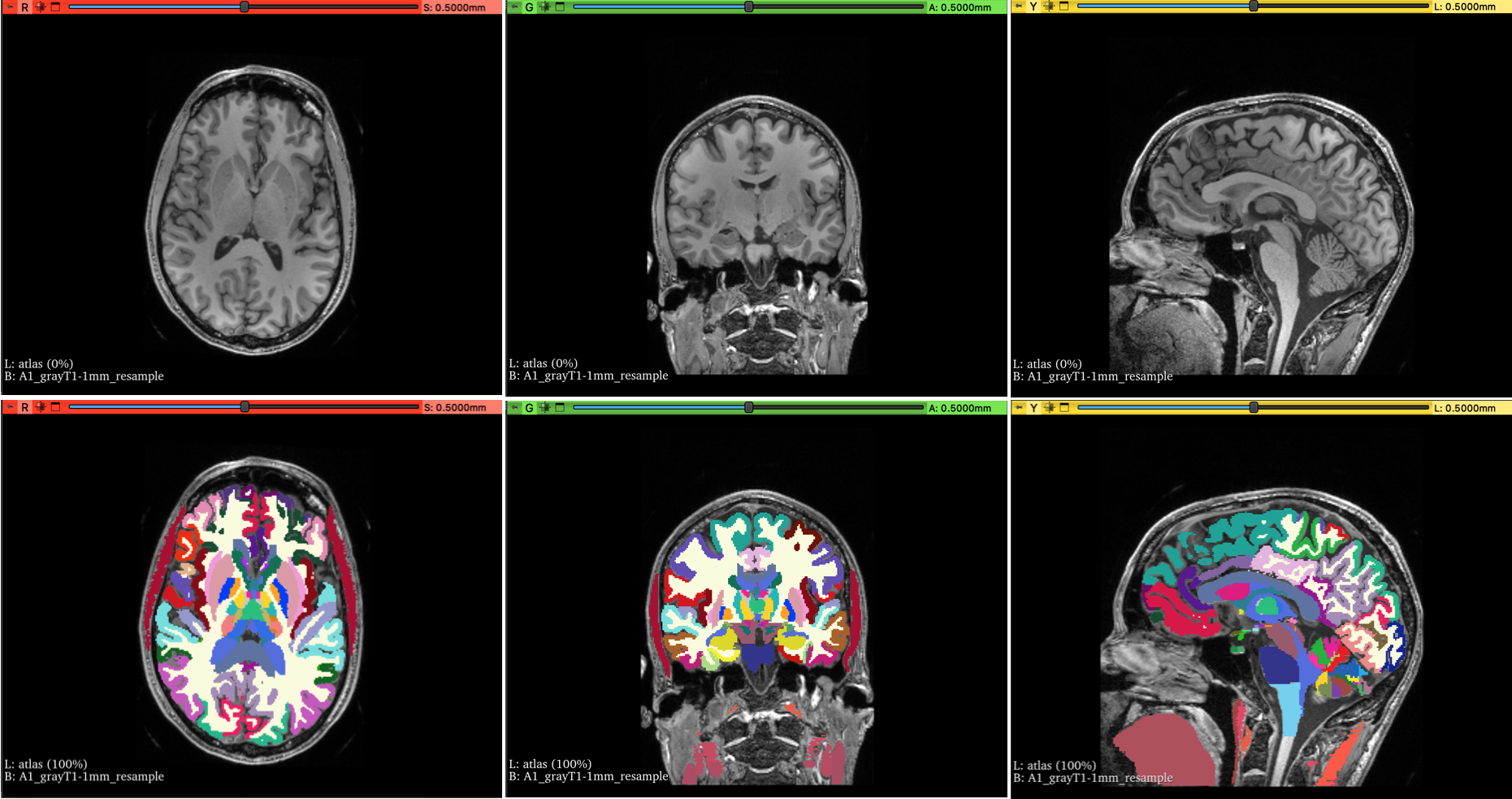}
    \caption{
    Open Anatomy Project’s SPL/NAC Brain Atlas. Top layer: T1-weighted MRI with 1 mm isotropic resolution. Bottom layer: SPL/NAC Brain Atlas label map overlayed with MRI.
    }
    \label{fig:fig1}
\end{figure*}

We developed the SlicerAtlasEditor software extension for 3D Slicer to edit Open Anatomy atlas label maps \citep{Huynh_Zwick}. The atlas editor allows users to remove, group and merge different anatomical labels. We grouped and merged all the anatomical structures which are comprised of white-matter and grey-matter for conductivity assignment. We also removed the skin and muscles from the label map as these were not used in the simulation. We then used our SlicerCBM software extension in 3D Slicer to generate labels for the cerebrospinal fluid (CSF), skull and skin \citep{Safdar_2023}. 

We used the original label map (Open Anatomy label map) for anatomical querying of simulation results and the modified label map (Property label map) for material property (conductivity) assignment.

\begin{figure*}
    \centering
    \includegraphics[width=\textwidth]{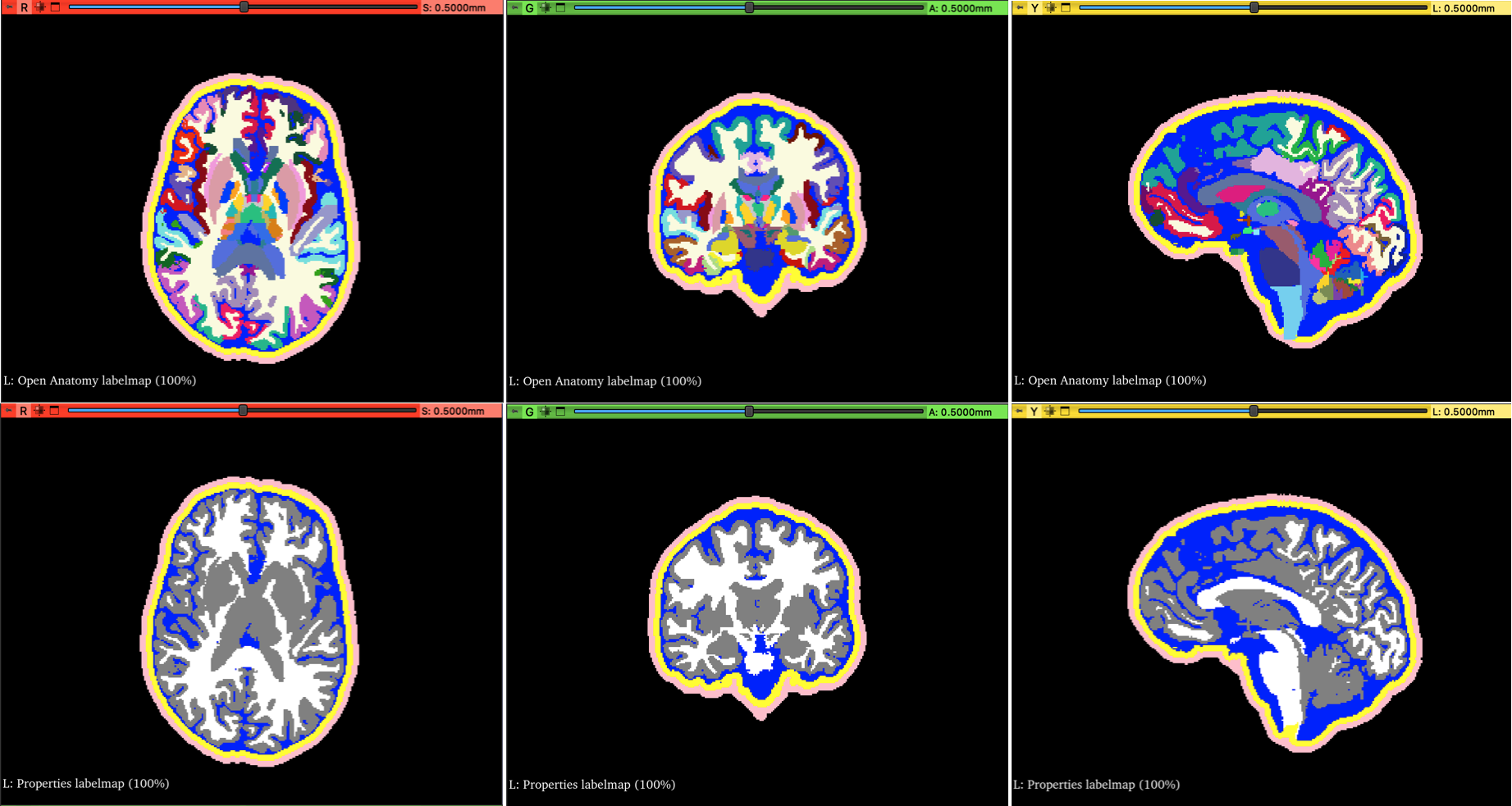}
    \caption{Brain label maps used for EEG forward problem. Top layer: Open Anatomy label map with additional CSF, skull and scalp labels. Bottom layer: Properties label map for conductivity assignment.}
    \label{fig:fig2}
\end{figure*}

\begin{table*}
  \centering
  \caption{Conductive compartments used in the EEG forward model.}
  \label{tab:conductivity}
  \begin{tabularx}{\textwidth}{@{}Xcc@{}}
    \toprule
    Compartment                & Conductivity (S/m) & References \\
    \midrule
    Scalp                      & 0.33               & \citet{Geddes_1967, Stok_1987} \\
    Skull                      & 0.012              & \citet{Hallez_2007,Gutierrez_2004} \\ %
    Cerebrospinal ﬂuid (CSF)   & 1.79               & \citet{Baumann_1997} \\
    Gray matter / White matter & 0.33               & \citet{Geddes_1967,Stok_1987} \\
    \bottomrule
  \end{tabularx}
\end{table*}

\subsection{Atlas-based hexahedral FE mesh creation}
The Open Anatomy label map is stored as an image volume in the NRRD format \citep{Halle_2017}. The label map is the output of a segmentation procedure that labels each voxel according to its segment (e.g., a certain type of tissue or structure). By converting the labelled voxels in the label map to 8-noded hexahedron elements, we can create a labelled hexahedral mesh with perfect quality elements. This is because the ideal hexahedral element is of the shape of a cube (e.g. shape of a voxel) \citep{Stimpson_2007}. We use the VTK library \citep{Schroeder_2006}, an open-source software for manipulating and displaying scientific data, to create the atlas-based  hexahedral FE mesh in VTK format. We do this by converting the label map image to VTK’s Unstructured Grid data object using an in-house python script. This data object is used for storing of topological and geometric properties and its associated attribute data and can be used as input to the FE solver. With the ability to store attribute data to the cells (elements) or points (nodes), we assigned scalar values which point to specific anatomical structures described in the colour map file. The colour map file is a lookup table representing discrete labels and colour information. This method creates a labelled FE mesh corresponding to the brain atlas. Since we have two types of label maps (Open Anatomy labels and Property labels), we generate the FE mesh with both labels as shown in Figure 3. Our FE mesh contains 2,318,585 hexahedral elements.

\begin{figure*}
    \centering
    \includegraphics[width=\textwidth]{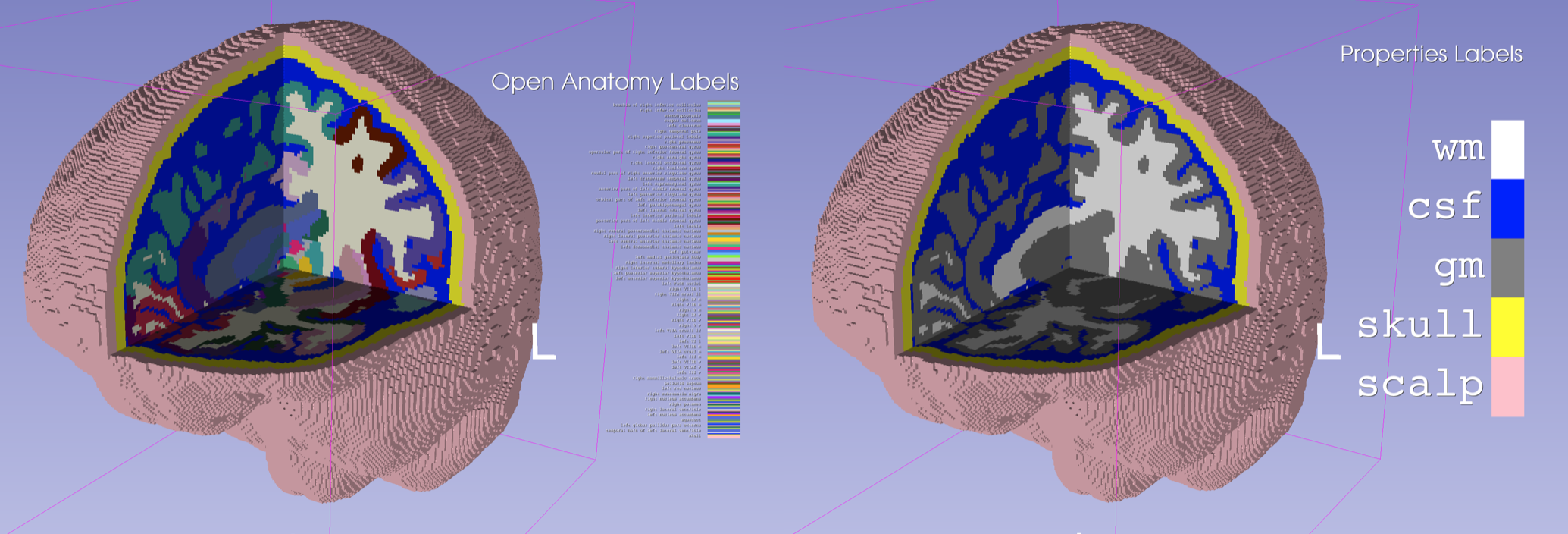}
    \caption{Atlas-based  hexahedral FE mesh. Left: FE mesh with Open Anatomy labels.  Right: FE mesh with Properties labels.}
    \label{fig:fig3}
\end{figure*}

\subsection{Solution of the EEG forward problem}

To generate the model for the EEG forward problem, we refer to our recent paper \citep{Zwick_2022} which solves a patient-specific EEG forward problem using the finite element method. The EEG forward problem involves predicting the electric potential within the brain given a predefined source. We define the dipole source in the medial temporal region of the brain, specifically the left middle temporal gyrus. This is to reproduce a common form of focal epilepsy called the ‘temporal lobe epilepsy’ (TLE) \citep{Javidan_2012}. The epileptic seizure onset source is typically modelled as a current dipole \citep{Hallez_2007}.

Following the literature, we approximate the physics of EEG by Poisson’s equation, which is a quasi-static approximation of Maxwell’s equation. For a spatial domain \(\Omega \in \mathbb{R}^3 \) with boundary \(\partial \Omega = \overline{\Gamma_D \cup \Gamma_N}\) and outward unit normal $n$, Poisson’s equation for the EEG forward problem can be written as:

\begin{align}
  \label{eq:poisson-eq}
  -\nabla \cdot (C (\nabla u)) &= f \text{ in } \Omega, \\
  \label{eq:poisson-nbc}
  \bm{n} \cdot (C (\nabla u))  &= g \text{ on } \Gamma_N, \\
  \label{eq:poisson-dbc}
  u &= h \text{ on } \Gamma_D,
\end{align}

where $u$ is the unknown scalar potential and $C$ is the tissue conductivity. 
We assign the conductivity to the atlas-based hexahedral FE mesh using the labelled elements acquired from the Property label map. We considered five tissue types: scalp, skull, cerebrospinal fluid (CSF), gray matter and white matter. The conductivity of the tissue compartments is given in Table 1. This method of tissue classification does not require any manual pre-segmentation as the atlas was used, greatly simplifying the model creation.

We solved the EEG forward problem using MFEM \citep{Anderson_2021} as described in our previous paper \citep{Zwick_2022}.

\section{Results}
\subsection{Visualisation of Simulation Results}

To display the simulation results with reference to the anatomical structures, we used ParaView, an open-source software for volume rendering and scientific visualization. Figure 4 show the electric potential field obtained from the dipole source with reference to the anatomical labels of the atlas. Our method offer users the ability to query results and their statistics by anatomical labels compared to previous methods that only used spatial coordinates, element number or node number. This provides additional information about the affected brain area, aiding researchers and medical professionals. We demonstrate this in Figure 5 by visualising the electric potential on the left middle temporal gyrus and its substructures.

\begin{figure*}
    \centering
    \includegraphics[width=\textwidth]{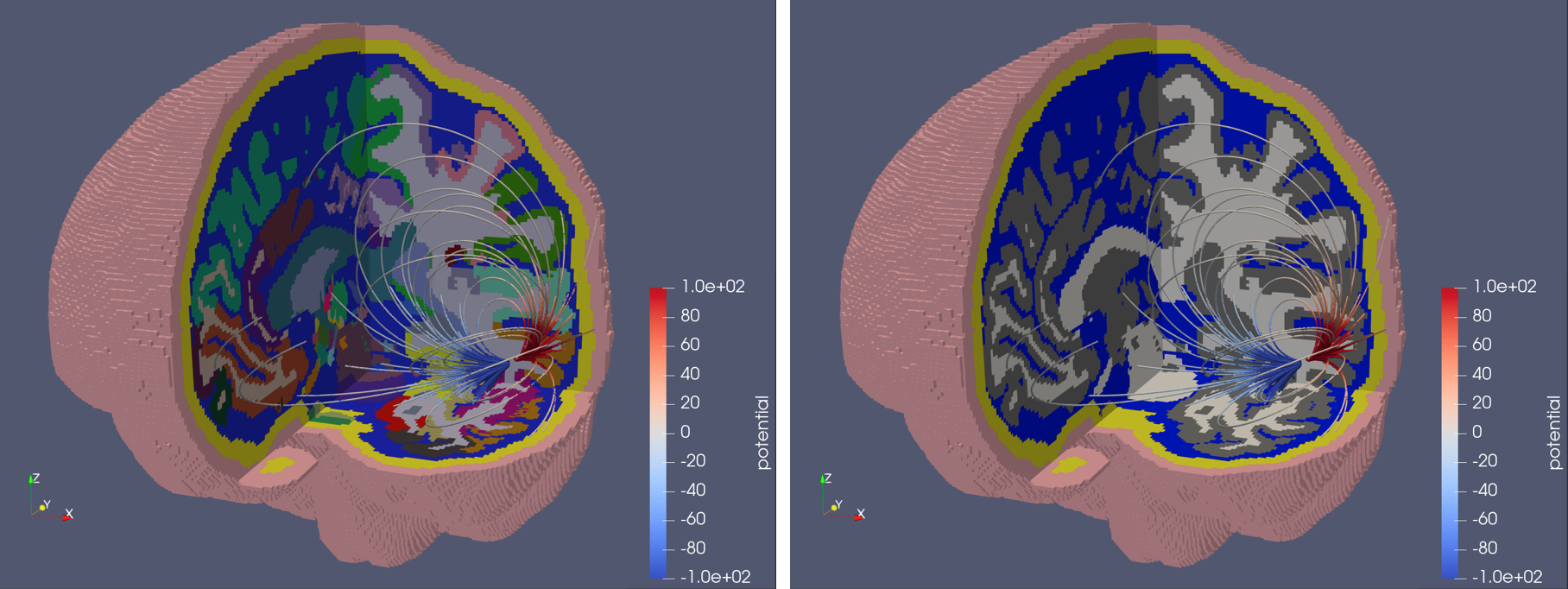}
    \caption{Cut-out of atlas-based FE mesh with streamlines of the electric field (in $\mu$V) generated by a current dipole source located inside the left middle temporal gyrus. The mesh elements are coloured according to (left) the anatomical labels and (right) the tissue conductivity compartments.}
    \label{fig:fig4}
\end{figure*}

\begin{figure*}
    \centering
    \includegraphics[width=\textwidth]{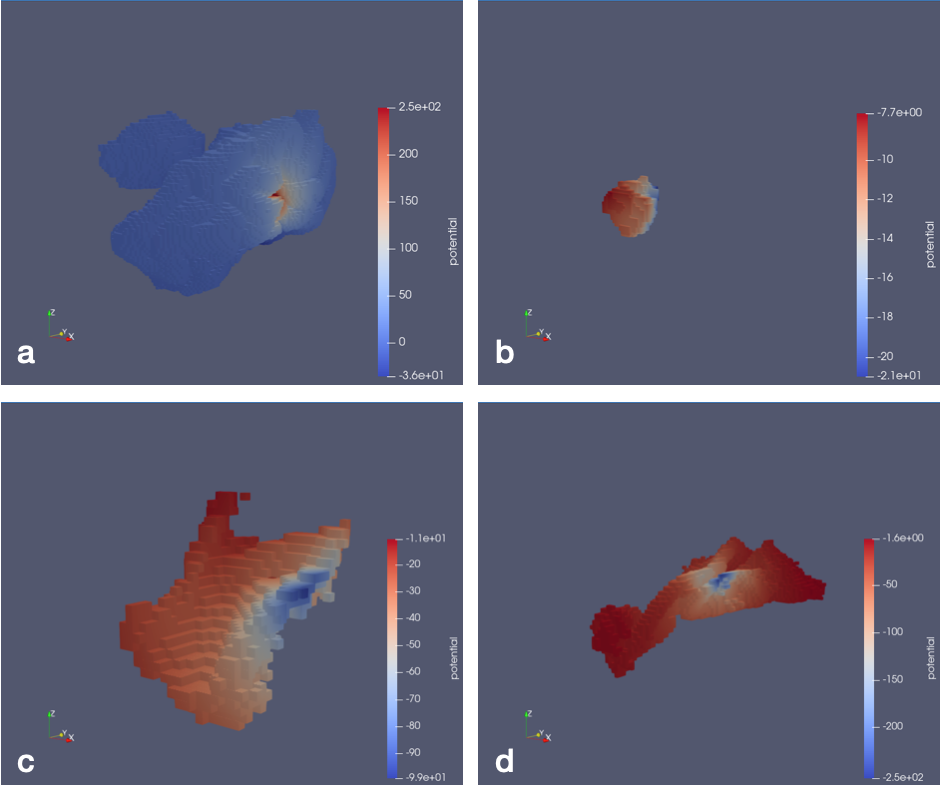}
    \caption{The electric potential (in $\mu$V) queried by the anatomical structure, namely the left middle temporal gyrus (a), left amygdala (b), left parahippocampal gyrus (c) and the left fusiform gyrus (d).}
    \label{fig:fig5}
\end{figure*}

\begin{table*}[ht]
  \centering
  \caption{Electrical potential (in $\mu$V) simulation results in the temporal lobe, amygdala, parahippocampus gyrus, and fusiform gyrus.}
  \label{tab:results}
  \begin{tabularx}{\textwidth}{@{}Xccc@{}}
    \toprule
    Anatomical Locations           & Maximum ($\mu$V) & Mean ($\mu$V) & Minimum ($\mu$V) \\
    \midrule
    Left middle temporal gyrus     & 21702.90	& 7.23 & -21702.90 \\
    Left amygdala                  & -7.66 & -12.36  & -22.24 \\ 
    Left parahippocampal gyrus     & -10.65 & -30.68 & -99.40 \\
    Left fusiform gyrus			 & -1.60 & -23.01  & 254.09 \\
    \bottomrule
  \end{tabularx}
\end{table*}

\subsection{Simulation results}
We demonstrate the benefits of using an atlas-based mesh by querying simulation results from specific regions of the brain, specifically the left middle temporal gyrus, where the dipole is located. We also query anatomical substructures, including the amygdala, parahippocampal gyrus, and fusiform gyrus, as these regions are typically implicated in the temporal lobe epilepsy seizure activity \citep{Corsellis_1970}. This approach holds significance since electro-physiological assessments primarily guide the decision-making process for surgical intervention in TLE patients \citep{Engel_1975}. 

We examine the maximum, minimum and mean potentials at the locations detailed in Table 2 above.

\section{Discussion}
In this paper, we have presented details of the creation of an atlas-based FE hexahedral mesh using the Open Anatomy brain atlas. The Open Anatomy Project currently stores and maintains several open atlases, including the brain, liver, inner ear, head and neck, knee, abdominal and thoracic atlas. We focus on utilising the SPL/NAC brain atlas; however, our work can be easily extended for other anatomical atlases. We provide a case-study simulation of an electrical field generated from a dipole located in the brain by solving the EEG forward problem. The two-way correspondence between anatomical structures in the atlas and the atlas-based mesh allows for easier FE mesh creation, material property (conductivity) assignment, and improved result visualisation and analysis. We avoid potential human errors from manual segmentation and reduce modelling efforts by using labelled structures from the Open Anatomy atlas. The SlicerAtlasEditor extension was used to create the ‘Properties' label map for conductivity assignments. The extension applies Boolean operations to group anatomical structures from the Open Anatomy brain atlas for material property assignment. Moreover, the atlas-based mesh enhances result visualisation and analysis by enabling querying of simulation results and their statistics based on anatomical structures. 

It should be noted that the meshes we have utilised in our study carry some limitations. Our mesh is of ‘high-quality’ due to the isotropic image resolution of the brain atlas, which results in creating ideal mesh elements. However, if the image resolution were to be anisotropic, the mesh elements would not be perfect, thereby leading to a decrease in mesh quality. To overcome this limitation, the images could potentially be resampled into an isotropic image. Further, our study utilises a mesh based on a single individual - a healthy 42-year-old male. This factor limits the extent of patient-specific applications we can build. Granted that there is a large variation in the brain's topology, geometry, and functional organisation across the population, it follows that our current mesh is most valuable for general applications that do not demand patient-specific geometries.

Nevertheless, our research aims to highlight the potential usefulness of creating FE hexahedral meshes directly from anatomical atlases. Future improvements to this approach might account for individual variation, making it potentially more beneficial for personalised medical applications.

\section*{Acknowledgments}

This research was carried out while the first author A.T. Huynh was in receipt of an “Australian Government Research Training Program Scholarship at The University of Western Australia”. Authors A. Wittek and K. Miller acknowledge funding from The University of Western Australia’s Research Income Growth Grant scheme for travel to Harvard University.

\bibliographystyle{elsarticle-harv}
\bibliography{references}

\nolinenumbers

\newpage
\listoffixmes

\end{document}